\documentclass[manuscript,times]{aastex631}

\accepted{September 1, 2022}
\submitjournal{AJ}

\shorttitle{X-rays from LTT 1445}
\shortauthors{Brown et al.}

\begin{document}

\title{X-ray Emission from the Exoplanet Hosting LTT 1445 Triple Star System}

\correspondingauthor{Alexander Brown}
\email{Alexander.Brown@colorado.edu}

\author[0000-0003-2631-3905]{Alexander Brown}
\affiliation{Center for Astrophysics and Space Astronomy, 
University of Colorado, 389 UCB, 
Boulder, CO 80309, USA}

\author[0000-0001-8499-2892]{Cynthia S. Froning}
\affiliation{McDonald Observatory, 
University of Texas at Austin, 
Austin, Texas 78712, USA}

\author[0000-0002-1176-3391]{Allison Youngblood}
\affiliation{Exoplanets and Stellar Astrophysics Lab, 
NASA Goddard Space Flight Center, 
Greenbelt, MD 20771, USA}

\author[0000-0002-1002-3674]{Kevin France}
\affiliation{Laboratory for Atmospheric and Space Physics, 
University of Colorado, 600 UCB, 
Boulder, CO 80309, USA}

\author[0000-0001-9667-9449]{David J. Wilson}
\affiliation{Laboratory for Atmospheric and Space Physics, 
University of Colorado, 600 UCB, 
Boulder, CO 80309, USA}

\author[0000-0002-7119-2543]{Girish Duvvuri}
\affiliation{Department of Astrophysical and Planetary Sciences, 
University of Colorado, Boulder, CO 80309, USA}
\affiliation{Center for Astrophysics and Space Astronomy, 
University of Colorado, 389 UCB, 
Boulder, CO 80309, USA}
\affiliation{Laboratory for Atmospheric and Space Physics, 
University of Colorado, 600 UCB, 
Boulder, CO 80309, USA}

\author{Yamila Miguel}
\affiliation{Leiden Observatory, P.O. Box 9500, 
2300 RA Leiden, The Netherlands}

\author[0000-0001-8274-6639]{Hannah Diamond-Lowe}
\affiliation{National Space Institute, Technical University of Denmark,  
Elektrovej 328, 2800 Kgs. Lyngby,  Denmark}

\begin{abstract}
JWST will be able to observe the atmospheres of rocky planets transiting nearby M dwarfs. 
The M dwarf triple star system LTT 1445, at a distance of 6.86 pc,  hosts some of the nearest 
rocky terrestrial planets. These planets most likely orbit the M 3.5V star LTT 1445A.
During a 28.6 ksec Chandra ACIS-S3 observation we have i) spatially resolved and detected all three 
stars in the LTT 1445 system,  ii) measured the X-ray luminosity of the individual stars, including 
LTT 1445A,  for the first time, iii) studied the flux variability of the X-ray sources and 
found strong variability from the A and C components, and iv) investigated how the 
coronal luminosities, temperatures and  volume emission measures vary at different activity levels.  
Combining these X-ray data  with upcoming HST ultraviolet observations will allow a  differential 
emission measure (DEM) estimation of the star's EUV spectrum, thereby facilitating modeling of 
the rocky planets' atmospheres.

\end{abstract}


\section{Introduction} \label{sec:intro}

Understanding what happens to rocky planets and their atmospheres in the habitable zones (HZs) of
low mass stars is currently one of the greatest astronomical challenges. The nearest Earth-mass
planets in the HZ orbit M dwarfs, and these are prime targets for spectroscopic atmospheric
characterization in the next decade (\citet{shields16}, \citet{shields19}).
M dwarfs are the most common type of star in the Galaxy, and $\geq$25 \% of them have planets 
orbiting in their habitable zones \citep{dressing15}.
Theoretical work shows that planets around M dwarfs could be habitable despite
their phase-locked orbits (\citet{joshi03}, \citet{ribas16}) and dynamic modeling
of transiting systems reveals that most systems permit stable orbits of Earth-mass planets
long enough for the development of life, i.e. $\geq$1.7 Gyr \citep{jones-sleep10}.
Ground-based surveys, such as MEarth \citep{nutzman08}  and the
SPECULOOS project \citep{reich13}, and space-based surveys like the Transiting Exoplanet
Survey Satellite (TESS; \citet{ricker15})
are finding and confirming nearby transiting planets orbiting M dwarfs whose atmospheres
should be observable with JWST \citep{luque19}. 
The best host stars for JWST atmospheric characterization will be nearby, slowly-rotating,
relatively inactive, mid-late M dwarfs with masses of 0.10-0.25 M$_{\odot}$ \citep{morley17}.

Over the past several years, the exoplanet community has recognized the importance of obtaining ultraviolet (UV)
and X-ray spectroscopy and time-series monitoring of M dwarfs to provide a comprehensive picture of their
energetic radiation environments. Surveys, such as Living with a Red Dwarf, MUSCLES, MegaMUSCLES,
HAZMAT, and FUMES (\citet{guinan16}, \citet{france16},  \citet{froning19}, \citet{wilson21}, \citet{Loyd_etal18}, 
  \citet{pineda21a}), are working to characterize the radiation fields of M stars as a function of
age, mass, and activity level. Such surveys are invaluable for understanding the
impact of the star's high energy irradiance on the atmospheric escape, chemical evolution, observability, and 
potential habitability of exoplanet companions (\citet{tilley19}, \citet{louca22}).  However, they have also
shown significant scatter in the relationships between stellar parameters and the high energy radiation, flare 
rates, and spectral energy distributions (e.g., see Fig. 3 of \citet{france18}). To interpret observations of
the atmospheres of potentially habitable planets in these systems, the optimum strategy is direct
panchromatic (X-ray/EUV/FUV/NUV/optical/IR) observations that can only be undertaken while the
space-based capabilities of HST and Chandra/XMM-Newton remain available.

\section{The LTT1445 Triple Star System} \label{sec:LTT1445}

In 2018 TESS discovered a new exoplanet of high potential for JWST atmospheric characterization 
-- a 2.87 M$_{\oplus}$ planet LTT 1445Ab in a 5.36-d orbit around the M dwarf,
LTT 1445A \citep{winters19}. Subsequently, \citet{winters22} discovered a second 1.54 M$_{\oplus}$ 
exoplanet in a 3.12 day orbit and improved estimates of the physical properties for both planets. 
At a distance of 6.864$\pm$0.001 pc \citep{GAIA20}, LTT 1445A is the nearest 
known transiting exoplanet system orbiting an M dwarf, which will substantially improve the ability 
of JWST to detect atmospheric features and direct thermal emission in a reasonable observing time. 
LTT 1445A is an approved JWST target for such an observation (program GO\#02708; PI Berta-Thompson).
Thus, as noted by \citet{winters19}, ``based on the known occurrence rates of planets orbiting M dwarfs,
it is unlikely that we will detect a small planet more favorable for atmospheric characterization.''
It is also bright enough for radial velocity follow-up to determine planetary masses and gravities
\citep{winters22}. 

LTT 1445 is a hierarchical triple star system with LTT 1445A separated by 7 arcseconds
from the tighter 36 year period visual binary BC.
The astrometric orbital motions of all three stars were measured by \citet{winters19}. 
LTT 1445A is an M3.5 V star with a mass of 0.257$\pm$0.014 M$_{\odot}$. 
The B and C components have masses of 0.215$\pm$0.014 M$_{\odot}$ 
and 0.161$\pm$0.014 M$_{\odot}$, which correspond to spectral types of 
M3.5 V and M4 V, when compared to masses and spectral types of other M dwarf 
exoplanet hosts (see e.g. \citet{pineda21b}). 
For comparison, \citet{reid04} assigned spectral types of M3, M3.5, and M4 to 
A, B, and C respectively.  \citet{winters19} estimated the bolometric luminosity of 
A to be 3.04 $\pm$ 0.12 $\times$ 10$^{31}$ ergs s$^{-1}$  . Scaling from the K magnitudes of the three 
stars (K = 6.50, 6.81, 7.33 for A, B, and C, respectively), the bolometric luminosities of the 
B and C components are 2.28  and 1.41 $\times$ 10$^{31}$ ergs s$^{-1}$. These 
values are entirely consistent with published values for M dwarfs with similar masses 
\citep{pineda21b}.

 The mass and radius measurements of LTT 1445Ab (2.87$\pm$0.26 M$_{\oplus}$;  
 1.305$\pm$0.066 R$_{\oplus}$) and LTT 1455Ac  (1.54$\pm$0.20 M$_{\oplus}$; 
 1.147$\pm$0.055 R$_{\oplus}$) indicate that both are likely to be terrestrial 
 with Earth-like compositions \citep{winters22}. 
 These planets lie interior to the habitable zone around the star with the cooler LTT 1445Ab 
 having an equilibrium 
temperature of 424$\pm$21 K, which is intermediate to other rocky planets such as GJ1132b and
TRAPPIST1-b. Although the known LTT 1445A planets are likely not habitable, they remain excellent targets for
detailed atmospheric studies with JWST; plus, given the frequency of multiple small planets around M dwarfs,
later TESS or ground-based radial velocity observations may reveal other planets in HZ orbits.
To properly interpret these and other studies of LTT 1445Ab and Ac, the high energy X-ray/UV context 
that governs exoplanet atmospheric properties must be established.

The LTT 1445 triple system was detected as a reasonably bright ROSAT PSPC source 
at a count rate of 0.25 cts s$^{-1}$\citep{boller16}, 
but it was unclear how the X-ray signal is distributed between the 3 stars.  Ground-based
spectra \citep{winters19} indicate that Balmer H$\alpha$ emission is detected from the BC binary,
suggesting that at least one of these stars is magnetically active.  
H$\alpha$ from the A component is in absorption, suggesting that this star is less likely to be the 
dominant X-ray emitter.

\section{Chandra Observations} \label{sec:obs}
The LTT 1445 system was observed by the Chandra X-ray Observatory on 2021 June 5 
using the ACIS-S3 back-illuminated detector in \onequarter subarray mode and VFAINT 
telemetry mode (ObsID 23377; PI Brown). The observation 
started at 01:41:32 UT and lasted 28.59 ksec.   The spatial resolution of ACIS is just under 
1 arcsecond (nominally 0\farcs84 at 50\% of the PSF peak but the CCD pixel size 
is only 0\farcs49) and is sufficient to resolve LTT 1445A from BC. 
The photon-counting capability of ACIS allows study of coronal flaring during
the observation and thus provides additional information on the high energy and overall 
SED variability for the stars.

\section{Data Analysis} \label{sec:analysis}
The ACIS-S data were analyzed using standard CIAO \citep{CIAO06} techniques 
on the standard pipeline-processed dataset obtained from the mission archive.
CIAO version 4.13 was used. Analysis was performed on data filtered to contain 
only events in the energy range 0.3 -10 keV. Events above 10 keV are unlikely to be of 
stellar origin and are usually due to high energy particles. ACIS now has minimal sensitivity
below 0.7 keV due to contaminant buildup on the optical blocking filters
\footnote{See  http:/cxc/harvard.edu/proposer/POG/html/chap6.html}. 
Energy filtering further reduced the already low detector background signal.
Source events were extracted using circular regions: for LTT 1445A and the combined 
BC source a  standard 2\farcs5  radius was used.  For exploring the individual contributions 
of the B and C components smaller 1\farcs0 and 0\farcs75 radii circles were used.

\subsection{Source Detection and Identification}\label{subsec:detect}

\begin{figure*}
\includegraphics[angle=0,scale=.90]{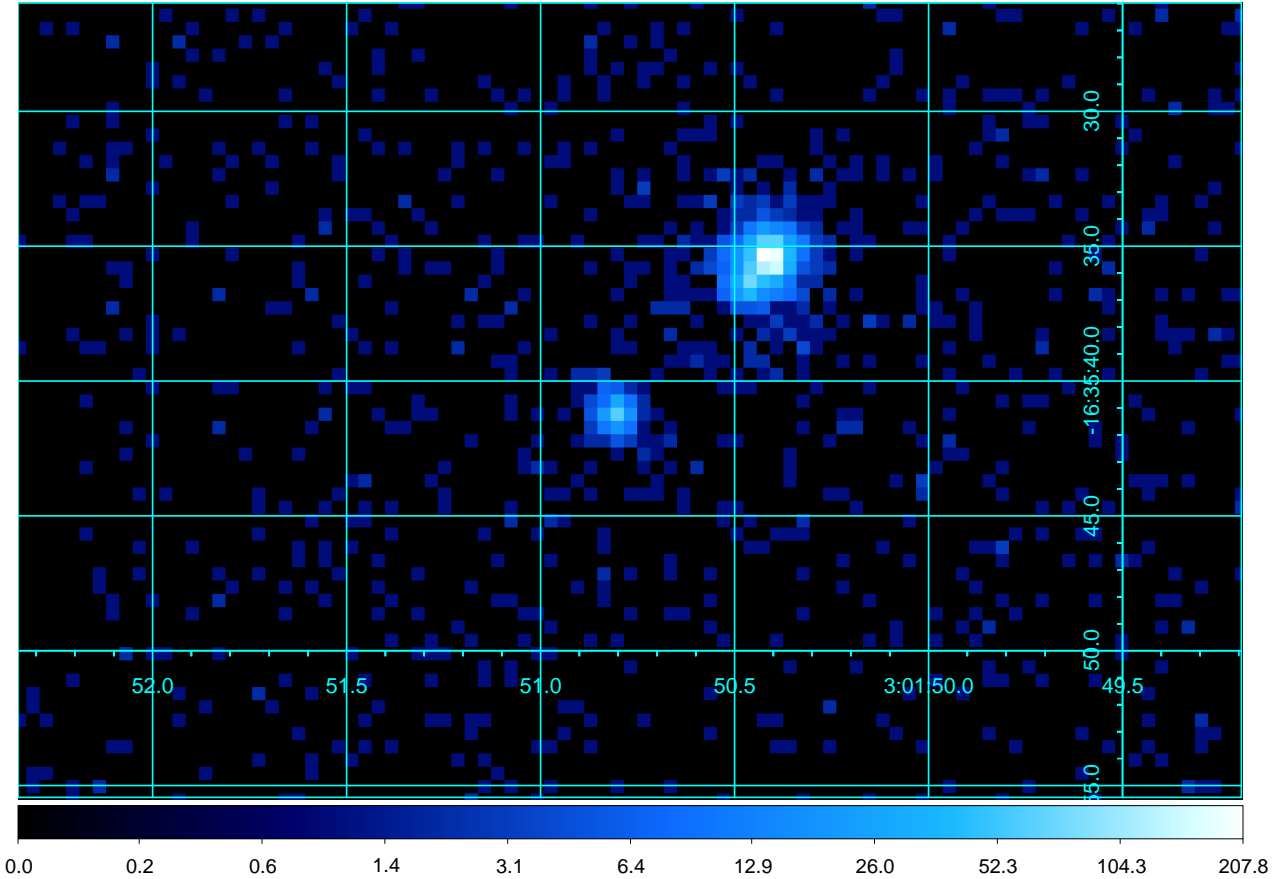}
\caption{The reconstructed Chandra ACIS S3 image of LTT 1445. North is up and east to the left.
The coordinate grid spacing is 5 arcseconds in declination and 0.5 seconds of time in right ascension.
The color wedge illustrates the integrated number  of counts per 0\farcs5 pixel. LTT 1445A is the fainter 
source at the center of the image with the brighter resolved BC binary to the NW.
\label{source_image3}}
\end{figure*}

Three X-ray sources are present in the ACIS image -- one well separated source at the position 
of LTT 1445A and a barely resolved  pair of sources coincident with LTT 1445BC (see Fig. \ref{source_image3}) 
The north-west source within the BC pair is by far the strongest X-ray emitter within the system.
LTT 1445A at a distance of 7 arcseconds was easily separated from BC. However, it was 
surprising that the LTT 1445BC binary was spatially resolved by ACIS, because the preliminary astrometric 
orbit estimated for the BC binary by \citet{winters19} implied that the two stars should have been far too close 
to be resolved in 2021 June. This discrepancy means that it is not possible from the Chandra data alone to 
associate the X-ray sources to the individual B and C components. The CIAO program {\it{wavedetect} }
was unable to separate the BC X-ray source but instead identified a single highly elliptical source 
encompassing both stars. The wavedetect centroid position for LTT 1445A agreed well with the 
proper-motion-corrected GAIA EDR-3 position \citep{GAIA16, GAIA20} with the offset between 
the two positions being only 0.8 arcseconds.

Fortunately, LTT 1445 was observed by the HST WFC3/UVIS instrument on 2021 Sep 26 and 29
using the F814W filter (program 16503; PI Winters; Obsids iejra2*, iejr03*). The passband of the 
F814W filter extends from 7000\AA\ to 9600\AA\ . These trailed images clearly show that the three
LTT 1445 components lie along the SE-NW orbit in the order A-B-C, with A being the optically 
brightest and C the faintest. Even though the stellar images are trailed, rather than point-like, 
it is still possible to estimate the separations well enough to remove any ambiguities in the X-ray data.
Thus, LTT 1445C is the brightest X-ray source in the system with $\sim$1,200 counts recorded 
from the full 28.6 ksec observation, while the B source has $\sim$250 counts and the A source 
is the weakest with $\sim$225 counts. In the X-ray image the A and C sources are separated 
by 8.25 arcseconds, the A and B sources by 7.0 arcseconds, and the B and C sources 
by 1.25 arcseconds -- these values are all in agreement with the optical imaging 
almost 4 months later. This location of C so far northwest of B is well beyond the 
maximum extent of the \citet{winters19} orbit, which has a maximum separation of 0.6 arcseconds 
in this direction. Additional optical data will undoubtedly resolve this discrepancy but any 
physical properties of this binary published in the 2019 paper should be treated with caution. 
However, none of the most critical stellar parameters in \citet{winters19} rely on the BC orbit.

\subsection{X-ray Variability}\label{subsec:variability}

\begin{figure*}[t!]
\includegraphics[angle=90,scale=.70]{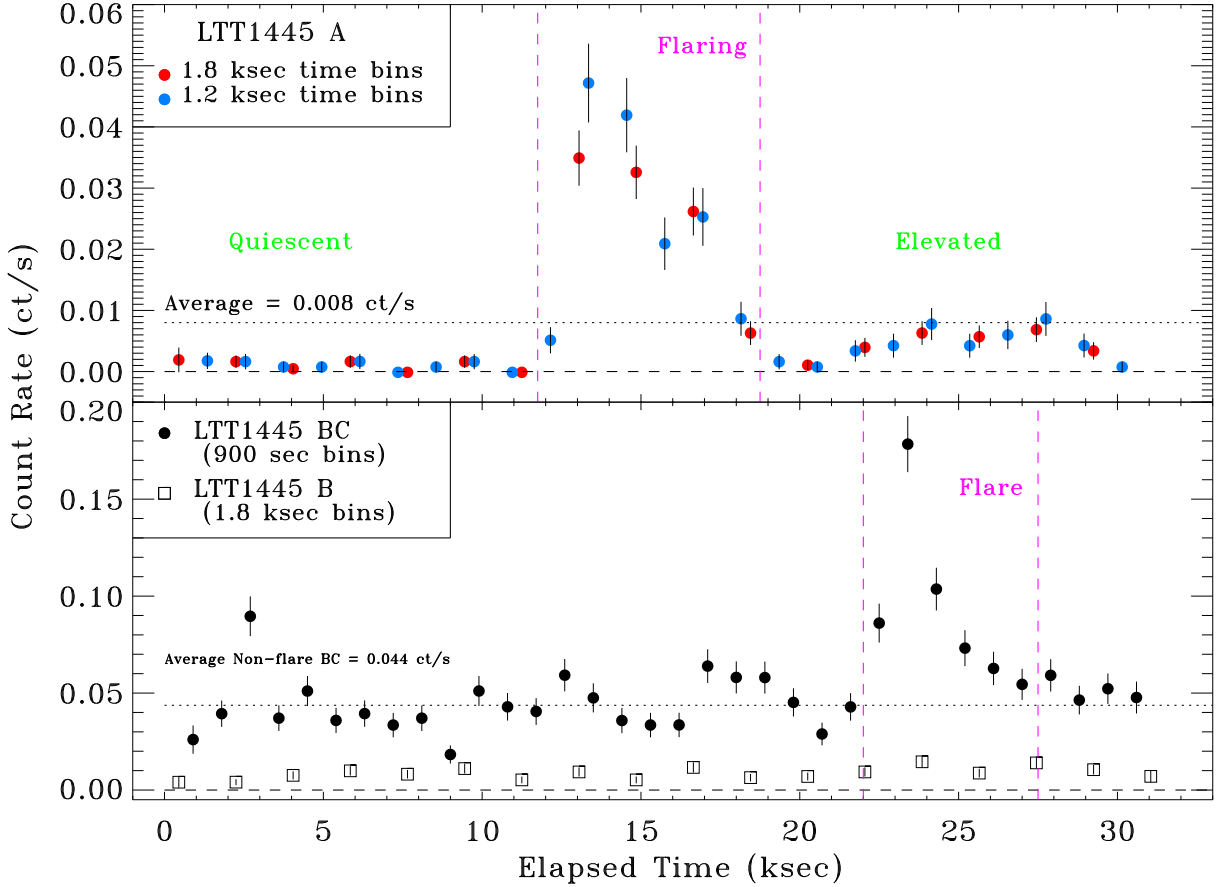}
\caption{The Chandra ACIS S3 light curves for the LTT 14445 X-ray sources. 
Upper panel: the light curve for LTT 1445A with different activity level segments separated by 
vertical dashed lines. Two different binning intervals (1.2 ksec (20 min) and 1.8 ksec (30 min) 
are plotted to better track the variability during the flare outburst.
Lower panel: The combined signal from the BC binary with 900 sec binning (filled dots), 
and the signal from the B component alone extracted using a small 1\farcs0 radius 
circular region and then scaled to recover the likely full PSF signal (open squares).
For the BC source the largest flare was also measured separately.
\label{lightcurves}}
\end{figure*}

After applying a barycentric correction to the ACIS data, X-ray light curves were constructed using the 
CIAO command dmextract using a variety of binning timescales and extraction region sizes. 
The light curve for LTT 1445A is shown in the upper panel of Fig. \ref{lightcurves} and shows that 
the X-ray emission from A is highly variable with a large flare present.
 The combined BC light curve and the scaled light curve from the 
PSF core of LTT 1445B are shown in the lower panel. 
The BC light curve shows significant variability throughout with the flaring originating from LTT 1445C.
Even though LTT 1445B lies on the PSF wings of the variable C source, 
the signal measured from the PSF core of B does not show significant variability aligned with 
major changes in the combined BC signal.

The LTT 1445A light curve has three distinct time intervals: an initial 12 ksec ``Quiescent'' segment with a very 
low count rate, then a 7 ksec long ``Flare'' outburst with multiple impulsive events, followed by just under 12 ksec 
with an  ``Elevated'' count rate at $\sim$10 times the quiescent level. At the end of the flare the count rate 
falls to the quiescent level before rising again. The reality of this variability is secure because the overall  
signal allows the LTT 1445A source region to be accurately positioned and the simultaneously recorded 
signal from the BC binary shows that ACIS was operating correctly. 
Data were extracted for the different time intervals and analyzed 
individually. Similarly, the largest flare from the BC source was also separated from the rest of those data.

\subsection{Spectral Analysis}\label{subsec:spectra}

Spectral fitting was performed using XSPEC V12.12 \citep{arnaud96, dorman03} to derive X-ray luminosities 
for all three sources and to estimate the coronal characteristic temperature and volume emission measure.  
The data for the A and BC components was split into different time intervals and the time variable plasma properties 
investigated. In most cases a single temperature (1-T) VAPEC model \citep{smith01} was sufficient 
to model the binned spectrum. However, for the ``Flare'' interval of the BC source  a 2-temperature model 
was heavily favored with a reduced $\chi^{2}$ of 1.26, compared to 2.2 for a 1-T fit. 
For all fits, the interstellar 
hydrogen column density was fixed at 1 $\times$ 10$^{19}$  cm$^{-2}$; this value consistently provided 
better $\chi^2$ values than lower interstellar columns. 
Generally, ACIS spectra cannot constrain low column densities well.
Initial sub-solar abundances were assumed as starting values based on results from M dwarf X-ray grating 
spectra analyses including \citet{raassen03}, \citet{vdBesselaar03}, \citet{guedel04},  and \citet{wargelin08}. 
Most importantly,  the Fe abundance was set to 0.4 relative to the XSPEC \citet{anders-grevesse} solar 
abundances. The Ca and Ni 
abundances were tied to that of Fe.  For C, N, O, Ne, Mg, Al, Si , S, and Ar  relative abundances of 1.0, 0.75, 
0.3, 0.8, 0.5, 0.5, 0.65, 0.4, and 0.55 were adopted. Most of these abundances have little influence in the fitting 
of ACIS CCD spectra in XSPEC. After model fitting with fixed abundances, the Fe abundance 
was allowed to vary to see if this provided a smaller reduced $\chi^2$ value.

\begin{deluxetable*}{ccccc}[t]
\tabletypesize{\footnotesize}
%\tabletypesize{\small}
\tablecaption{Coronal X-ray Properties  for LTT 1445A \label{table_xspecfits_A}}
\tablewidth{900pt}
\tablehead{
\colhead{ }  &  \colhead{A - Full Dataset} &\colhead{A - Flare} & \colhead{A - Elevated} & \colhead{A - Quiescent} }
\startdata 
Exposure (ks)&28.6 & 6.66 & 11.6 & 12.2 \\
Source Counts (ct) & 228  & 177 & 46 & 4.9  \\
Count Rate (ct ks$^{-1}$) & 8.0$\pm$0.5  & 26.6$\pm$2.0 & 4.0$\pm$0.6  & 0.4$\pm$0.2 \\
f$_{X}$ (10$^{-13}$ ergs cm$^{-2}$ s$^{-1}$)  & 1.52$\pm$0.10 & 3.61$\pm$0.27 &0.66$\pm$0.10&0.066$\pm$0.033  \\
log L$_{X}$    & 26.93$\pm$0.03 & 27.31$\pm$0.03 & 26.57$\pm$0.07 &  $25.57_{-0.30}^{+0.18}$ \\
log L$_{X}$/L$_{bol}$   &  -4.55$\pm$0.04  &  -4.17$\pm$0.04 &   -4.91$\pm$0.07 &  $-5.90_{-0.3}^{+0.2}$   \\
kT$_{1}$ (keV)   &  0.63$\pm$0.08 & 1.02$\pm$0.10 & 0.59$\pm$0.29  &   \nodata \\
norm1 ($10^{-4}$) & 1.16$\pm$0.36 & 2.24$\pm$0.80 & 0.38$\pm$ 0.34  &  \nodata  \\
VEM$_{1}$ (10$^{49}$ cm$^{3}$) & 0.52$\pm$0.16 & 1.00$\pm$0.36 & 0.17$\pm$0.15 &  \nodata  \\
Fe Abundance & 0.20$\pm$0.07 & 0.42$\pm$0.14 & Fixed  0.4 solar & \nodata \\
Red. $\chi^{2}$ & 0.87 & 0.88 & 1.7  &  \nodata \\
\enddata
\end{deluxetable*}

The X-ray properties for LTT 1445A are listed in Table \ref{table_xspecfits_A}. The X-ray fluxes and luminosities 
listed are for the 0.3-10.0 keV energy range. XSPEC fits were performed for the 
full dataset and the ``Flare'' and ``Elevated'' portions of the light curve. The initial ``Quiescent'' interval contains 
too few counts for spectral fitting but the X-ray flux and luminosity were estimated assuming that the coronal 
temperature was similar to that during the ``Elevated'' time interval. 
The range in the X-ray emission from LTT 1445A is large with the count rate going from 0.4 counts/ksec in 
quiescence to 47.2 counts/ksec at the flare peak --- thus, varying by a factor of $\sim$ 100. However, the 
quiescent X-ray luminosity is at a low level of 3.7 $\times$ 10$^{25}$ ergs s$^{-1}$ .

\begin{deluxetable*}{cccccc}[t]
\tabletypesize{\footnotesize}
%\tabletypesize{\small}
\tablecaption{Coronal X-ray Properties  for LTT 1445BC \label{table_xspecfits_BC}}
\tablewidth{900pt}
\tablehead{
\colhead{ }  & \colhead{CB - Full Dataset} &    \colhead{CB - Flare} &   \colhead{CB - Non-Flare} &    \colhead{C}  &    \colhead{B}  }
\startdata 
Exposure (ks)& 28.6 & 5.2 & 23.4 & 28.6 & 28.6 \\
Source Counts (ct) & 1,468 & 470 &  997 &  1218 &  250\\
Count Rate (ct ks$^{-1}$) & 51.3$\pm$1.3 & 89.9$\pm$4.2  & 42.7$\pm$1.4 & 42.5$\pm$1.2   &  8.7$\pm$0.6 \\
f$_{X}$ (10$^{-13}$ ergs cm$^{-2}$ s$^{-1}$)  & 9.4$\pm$0.2  & 20.2$\pm$0.9 & 7.3$\pm$0.2  & 7.3$\pm$0.2  & 1.7$\pm$0.1  \\
$\log$ L$_{X}$    & $27.72\pm0.02$ & $28.06\pm0.02$ & $27.61\pm0.01$ &  $27.62\pm0.01$ &  $26.97\pm0.03$  \\
log L$_{X}$/L$_{bol}$   &  -3.84$\pm$0.062 &  -3.51$\pm$0.02 &   -3.95$\pm$0.01 &  -3.53$\pm$0.01 &  -4.39$\pm$0.03   \\
kT$_{1}$ (keV)   & 0.77$\pm$0.03   & 0.38$\pm$0.10  & 0.76$\pm$0.03 &   &   \\
norm1 ($10^{-4}$) &  4.1$\pm$1.7 &   7.3$\pm$4.5  & 4.1$\pm$0.3  &   &   \\
VEM$_{1}$ (10$^{49}$ cm$^{3}$) & 1.86$\pm$0.71 & 3.27$\pm$2.02 & 1.84$\pm$0.13 \\
kT$_{2}$ (keV)      &  \nodata   &   2.2$\pm$0.5&  \nodata  &   &  \\
norm2 (10$^{-4}$)   &  \nodata  &  5.0$\pm$1.5  & \nodata   &   &   \\
VEM$_{2}$ (10$^{49}$ cm$^{3}$) & \nodata   & $2.23\pm0.66$ &  \nodata  \\
Red. $\chi^{2}$ & 1.23 \tablenotemark{a}& 1.26 \tablenotemark{b} & 1.16 \tablenotemark{c} &   &   \\
\enddata
\tablenotetext{a}{Fitted abundances: Fe: 0.38$\pm$0.08, Si: 0.50$\pm$0.16, Ne: 1.26$\pm$0.56}
\tablenotetext{b}{Fitted abundances: Fe: 0.67$\pm$0.29}
\tablenotetext{c}{Fe abundance fixed to 0.4 solar}
\end{deluxetable*}

The X-ray properties for LTT 1445BC are listed in Table \ref{table_xspecfits_BC}. XSPEC fits were performed for the 
full dataset and both ``Flare'' and ``Nonflare'' portions of the light curve. However, given the continuous variability 
of the BC source, both these segments almost certainly contain numerous flares.  For the combined CB measurements, 
the summed bolometric luminosities are used to estimate log L$_{X}$/L$_{bol}$.  The properties of the individual 
C and B sources are also estimated, based on the count levels at the peak of the PSFs.

\begin{figure*}[t!]
\includegraphics[trim= 50mm 0mm 40mm 0mm, angle=-90,scale=.66]{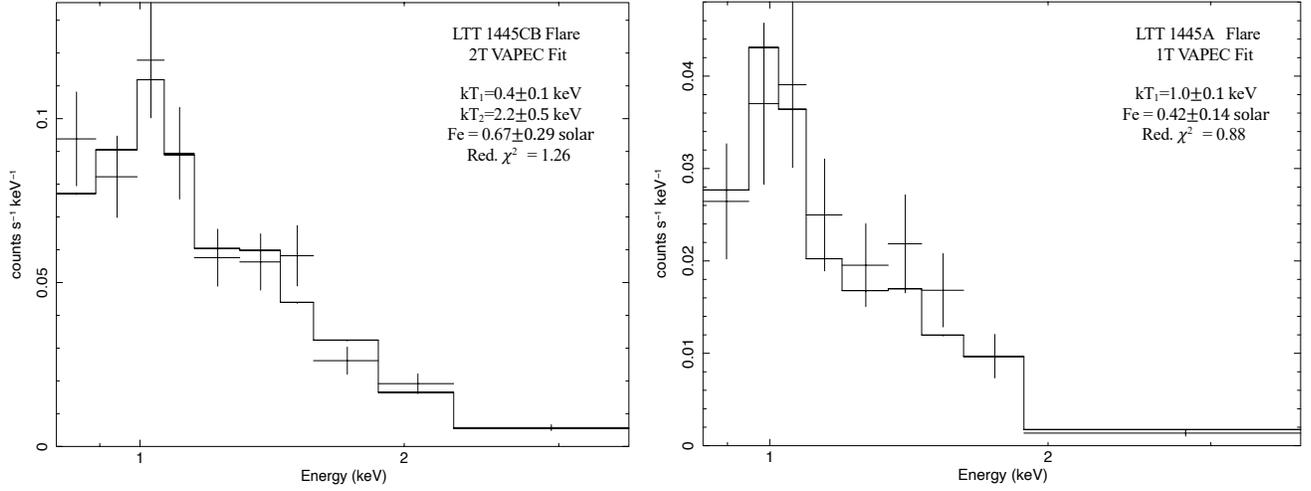}
\caption{Examples of XSPEC spectral fitting to Chandra CCD-resolution spectra. 
These figures demonstrate the fitting framework within XSPEC.
 The spectral data with associated error bars are overlaid on the final model shown as a histogram.
Right panel: a one-temperature (1-T) fit to the flare spectrum of LTT 1445A grouped with 17 events per channel.
Left panel: a two-temperature (2-T) fit to the flare spectrum of LTT 1445 CB grouped with 40 events per channel.
\label{spectralfits}}
\end{figure*}

Examples of the spectral fitting within XSPEC are shown in Figure \ref{spectralfits}. 
The binning of the spectra is a delicate balance between maintaining sufficient events in each spectral channel 
while providing enough valid channels to solve for the number of fit parameters. 
The presence of flaring plasma, clearly seen in the light curves, is confirmed by the  coronal characteristic temperatures 
measured by the 1-T fits, which are in the range 0.6-0.8 keV (7-9 MK). Temperatures at this level are hotter than typical 
quiescent coronal temperatures of low activity M dwarfs  (0.3-0.4 keV; 3.5-5 MK -- \citet{brown22}) and indicate 
a significant presence of flaring and cooling post-flare 
plasma in the coronae of LTT 1445 A and C. This is emphasized by the 2-T fit to the LTT 1445CB ``Flare'', where 
comparable amounts of cool 0.38 keV (4.5 MK) and very hot 2.2 keV (25 MK) plasma are present. 
The corresponding 1-T coronal temperature was 1.05$\pm$0.07 keV.
Using a 1-T model can often result in a parameterization favoring intermediate values when a wide range of plasma 
conditions are present.

\section{Discussion} \label{sec:discuss}

\subsection{X-ray Activity Levels}\label{subsec:Lx}

\begin{figure*}
\includegraphics[angle=90,scale=.65]{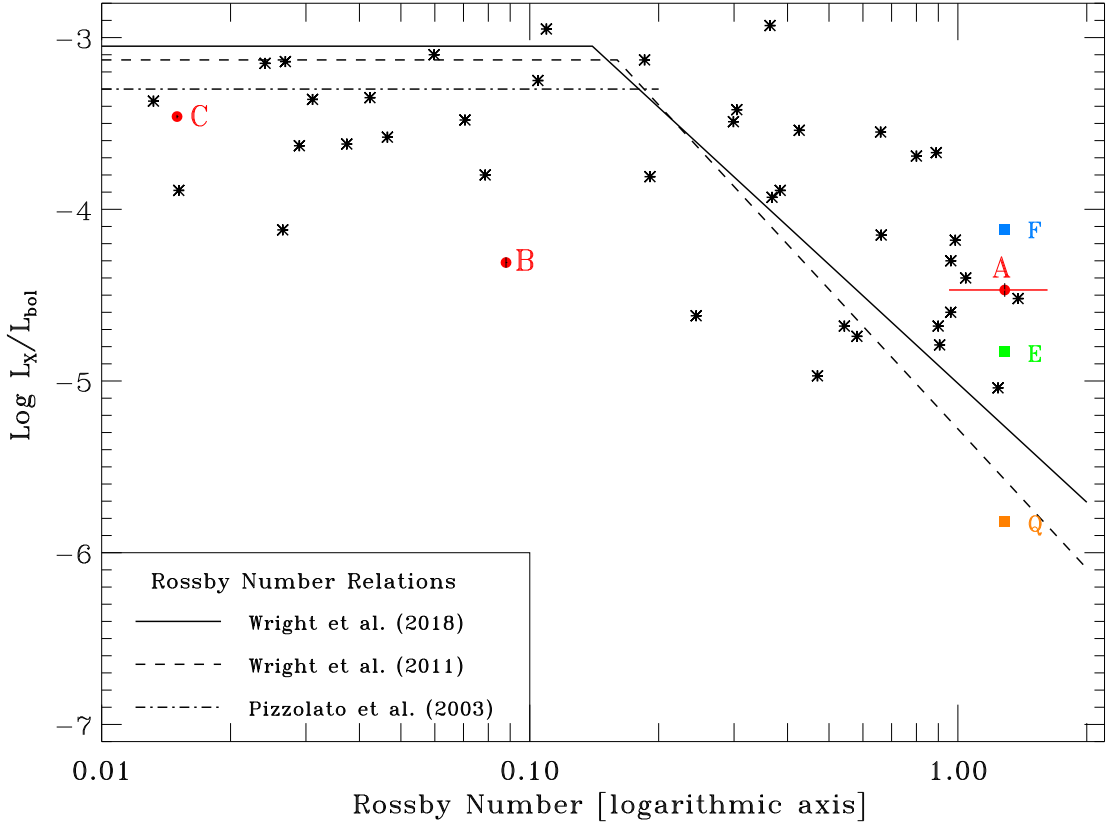}
\caption{X-ray to bolometric luminosity ratios for the LTT 1445 
X-ray sources as a function of Rossby number. Mean X-ray emission for each star is
shown as red circles. The different activity levels of LTT 1445A are shown as filled squares:
quiescent (Q - orange), elevated (E - green), and flare (F - blue). The nearby M dwarf sample 
included in \citet{wright18} and \citet{wright11}  are shown as black asterisks. The empirical relationships 
between the X-ray to bolometric luminosity ratio and Rossby number (rotation period/convective 
turnover timescale) derived by \citet{wright18} and \citet{wright11} from larger samples of F-M type dwarfs 
are plotted as solid and dashed lines, respectively. The M dwarf saturated activity level of \citet{pizzolato03} is
shown as a dot-dash line.
\label{fig:luminosities_rossby}}
\end{figure*}

The measured X-ray luminosities for the three stars in the LTT 1445 system can be 
compared to the behavior of the wider solar neighborhood M dwarf population. 
Stellar rotation is the dominant factor controlling magnetic activity on M dwarfs 
(\citet{noyes84}, \citet{pizzolato03}, \citet{west15}, \citet{magaudda20}).
Initially, young rapidly rotating M dwarfs show saturated coronal emission with 
log L$_X$/L$_{bol}$ $\simeq$ -3 . Then, as rotation slows, the 
X-ray luminosity decreases. 
\citet{wright18} and \citet{wright11} provide a detailed study of the relationship 
between X-ray luminosity in the ROSAT 0.1-2.4 keV energy range and stellar 
rotation for a wide range of dwarf stars with F-G-K-M spectral types.
They showed that the tightest correlations involved the Rossby number 
(R$_0$ = P$_{rot}$/$\tau$, where $\tau$ is the convective turnover time),
rather than the rotation period (P$_{rot}$) alone. The relationships discussed by 
\citet{wright18} and \citet{wright11} are shown in Fig \ref{fig:luminosities_rossby}. 
The declines from saturated emission start at R$_0$ = 0.14 and 0.16.
As a guide to the range of coronal activity detected, the sample of nearby M dwarfs included in 
these papers are plotted on Fig. \ref{fig:luminosities_rossby}. The Rossby numbers for 
all these stars were recalculated using Eqn. 5 of \citet{wright18}.

\citet{winters22} estimated the rotation periods of the A, B, and C components to be 
85$\pm$22, 6.7 and 1.4 days, respectively, which correspond to Rossby numbers of 
1.287$\pm$0.333, 0.088, and 0.015. Therefore, LTT 1445 B and C should be fast 
rotators and show saturated activity.
The observed X-ray to bolometric luminosity ratios for the three LTT 1445 stars 
 are plotted in Fig \ref{fig:luminosities_rossby}, as well as those for different 
 activity levels shown by LTT 1445 A. These X-ray luminosities were converted from the 
 Chandra 0.3-10.0 keV energy range to the ROSAT band using PIMMS \footnote{The 
  Portable Interactive Multi-Mission Simulator, http:/cxc/harvard.edu/toolkit/pimms/jsp } 
  and the spectral properties listed in Tables \ref{table_xspecfits_A} and \ref{table_xspecfits_BC}.
 Converting from the Chandra to ROSAT energy band increases the X-ray luminosity by a 
 factor that is temperature dependent (29.7\% increase at 0.4 keV, 20.6\% increase at 0.6 keV,
 17.9\% increase at 0.8 keV, and 11.7\% at 1.1 keV).
 Interestingly, the large observed range in the X-ray emission from LTT 1445A 
 covers the wide spread in log L$_X$/L$_{bol}$ seen for other M dwarfs, which provides  
further support to the suggestion that much of that spread is due to flaring variability. 
 While the L$_X$/L$_{bol}$ for LTT 1445C lies 0.5 dex below the saturation limit 
 of \citet{wright18}, it is less than 0.2 dex below that of \citet{pizzolato03}.

 While the A and C components have X-ray emission 
 in agreement with the activity-R$_0$ relationships, the X-ray emission from the B  
component is a factor of $\sim$20 below the \citet{wright18} saturated level for a star with a 6.7 day 
rotational period. The observed L$_X$/L$_{bol}$ for LTT 1445B corresponds to a 
rotation period of $\sim$45 days on the relationship of \citet{wright18}. 
However, many other M dwarfs with similar Rossby numbers also lie significantly below 
the nominal saturated level and it is unclear whether LTT 1445B was observed during 
quiescence and future observations will show brighter coronal emission or if it is indeed 
 a more slowly rotating star.

\subsection{Flares and Coronal Variability}\label{subsec:flares}
An important factor influencing the atmospheric stability and photochemistry of extrasolar
planets is temporal variability of the energetic incident radiation. Impulsive magnetic flares
can be related to energetic particle ejections. \citet{segura10} showed
that energetic particle deposition into the atmosphere of an Earth-like planet during large M dwarf
flares can lead to significant atmospheric ozone depletion ($\geq$ 90\% for extreme flares).
This alters the atmospheric chemistry and increases the penetration depth of UV photons that
are damaging to surface life \citep{tilley19}. Most M stars exhibit significant X-ray/UV variability, 
even for older systems (see e.g \citet{france20}), as evidenced by the many
flares seen in MUSCLES  observations \citep{Loyd_etal18}.
These flares can be comparable to or exceed the quiescent flux of the star (\citet{Loyd_etal18};
\citet{froning19}). On the other hand, the intense
flare flux may be needed to spur the formation of life, as the low quiescent NUV flux
from M stars may be insufficient to drive UV-sensitive prebiotic nucleotide synthesis
(\citet{ranjan17}; \citet{rimmer18}).

Even though the X-ray luminosity of LTT 1445A is weak, the range of variability is very large.
This implies that the high energy irradiance on its planets will be highly variable and lead to 
stochastic atmospheric changes. Our 28.6 ksec view of the X-ray emission from LTT 1445A 
provides only a limited window into the flaring duty cycle. Fortunately, a longer 50 ksec ACIS-S  
observation is scheduled in the second half of 2022 (PI: Predehl; ObsID: 25993), which 
should provide improved insights into the temporal variability of X-ray emission from this star.

\subsection{Stellar Spectral Energy Distribution}\label{subsec:SED}

Knowledge of the complete stellar SED is vital when modeling the surface and atmospheric conditions
of any exoplanet and X-ray observations alone cannot provide a complete description of the SED. 
Optical/IR photons are the dominant overall heating source that determines the
location of the habitable zone around the star.
Ultraviolet stellar radiation drives atmospheric heating and chemistry on Earth-like planets,
while EUV/X-ray radiation forces thermospheric heating and atmospheric escape and erosion.
Most of the stellar SED can be observed directly except for the extreme-ultraviolet
(EUV; 100 $\leq \lambda \leq $ 911 \AA) which must be modeled.
EUV photons from the central star are an important source of atmospheric heating and ionization on
all types of extrasolar planets. For terrestrial atmospheres, increasing the EUV flux to levels
estimated for the young Sun ($\sim$1 Gyr; \citet{ayres97}) can increase the temperature of the
thermosphere by a factor of $\geq$10 \citep{tian08}, potentially causing significant and
rapid atmospheric mass-loss. Estimates of the incident EUV flux are therefore important to
the long-term stability of an exoplanetary atmosphere;
however, direct measurement of the complete EUV irradiance spectrum is extremely difficult 
because of a lack of EUV space observatories and because attenuation by interstellar hydrogen
prevents detailed characterization for most stars except the Sun.
Stellar EUV emission is a combination of emission from both transition region ($\sim$10$^5$ K) plasma
and coronal ($\geq$10$^6$ K) plasma. Other emission lines from these two thermal regions can be
observed directly in the FUV and X-ray spectral regions. Therefore, the most accurate and reliable method
to calculate the EUV radiation field is via the combined differential emission measure (DEM) analysis
of FUV and X-ray spectra \citep{louden17}.

We intend to model the CCD-resolution spectra measured by Chandra in combination with planned 
future HST FUV/NUV spectra (Program 16701, PI Youngblood; Program 16722, PI Diamond-Lowe) 
using the MCMC DEM modeling 
technique \citep{duvvuri21}, which is currently being used to analyze the SEDs of other M dwarf exoplanet 
host stars. This should provide full irradiance spectra that can be used in atmospheric modeling of the 
exoplanets orbiting LTT 1445A.

\section{Conclusions} \label{sec:conclude}

Chandra ACIS-S observations have detected and resolved the coronal X-ray emission from all three stars 
in the LTT 1445 system, including from the exoplanet host star LTT 1445A.  The dominant X-ray emitter 
in the system is the lowest mass star LTT 1445C, which is the most likely source of 1.4 day period 
optical variability seen in TESS light curves. The Chandra observations of LTT 1445A provide further 
evidence that the coronal emission of older, slowly-rotating M dwarfs can be highly variable and the 
coronal emission already seen from LTT 1445A 
suggests that this star may have interesting influences on the atmospheres of  its exoplanets.
During the observed flare the X-ray luminosity is only a factor of 10 below the 
saturated activity level expected from young, very active stars, even though LTT 1445A is 
thought to be slowly rotating. More extensive studies of its X-ray activity should provide a clearer 
picture of the range of coronal emission from all three mid-M dwarfs in the LTT 1445 system. 

\begin{acknowledgments}
We thank the referee for useful comments and suggested additions to this paper. 
This research has used data obtained by the Chandra X-ray Observatory and software provided by 
the Chandra X-ray Center (CXC) in the CIAO application package. 
This work was supported by Chandra grant GO1-22005X to the University of Colorado.
An HST/WFC3 optical image of the LTT 1445 system was obtained from the Mikulski Archive for Space 
Telescopes (MAST) at the Space Telescope Science Institute, which is operated by the Association 
of Universities for Research in Astronomy Inc., under NASA contract NAS5-26555.

We acknowledge the invaluable resources of NASA's Astrophysics Data System and the SIMBAD/VIZIER 
databases operated by CDS, Strasbourg, France \citep{wenger00}.

\end{acknowledgments}

%\vspace{5mm}
\facilities{CXO (ACIS), HST (WFC3)}

%\software{astropy \citep{2013A&A...558A..33A},  
 %         Cloudy \citep{2013RMxAA..49..137F}, 
  %        SExtractor \citep{1996A&AS..117..393B}
   %       }
\software{CIAO \citep{CIAO06}, XSPEC \citep{arnaud96}, IDL (Ver. 8.8; Excelis Visual Information Solutions, Boulder, CO)}

%% For this sample we use BibTeX plus aasjournals.bst to generate the
%% the bibliography. The sample63.bib file was populated from ADS. To
%% get the citations to show in the compiled file do the following:
%%
%% pdflatex sample63.tex
%% bibtext sample63
%% pdflatex sample63.tex
%% pdflatex sample63.tex

%\bibliography{sample63}{}
%\bibliographystyle{aasjournal}

%% This command is needed to show the entire author+affiliation list when
%% the collaboration and author truncation commands are used.  It has to
%% go at the end of the manuscript.
%\allauthors

%% Include this line if you are using the \added, \replaced, \deleted
%% commands to see a summary list of all changes at the end of the article.
%\listofchanges

\end{document}